\documentstyle[12pt,aaspp4,epsf]{article}

\begin{document}

\def\intldate{\number\day\space\ifcase\month\or
January\or February\or March\or April\or May\or June\or
July\or August\or September\or October\or November\or December\fi
\space\number\year}

% ----------------------------------------------------------------------------

\def \event	{{MACHO~97-BLG-41}}
\def \ncite  	#1{#1\ --{\it get ref}}
\def \deg    	{$^{\circ}$}
\def \etal   	{{et al.\thinspace}}
\def \eg     	{{e.g.,}}
\def \cf     	{{cf.}}
\def \ie     	{{i.e.,}}
\def \hub    	{$H_{\hbox{\rm 0}}$}
% \hunits and \kms have been fixed here
\def \hunits 	{\mbox{${\rm km\,s^{-1}\,{\rm Mpc}^{-1}}$}}
\def\kms	{\mbox{${\rm km}\,{\rm s}^{-1}$}}
\def \sec    	{$^{s}$}
\def \arcsecpoint {{$^{''}\mkern-5mu.$}}
\def \asec	{{\arcsecpoint}}
\def \dophot	{D{\sc o}PHOT}
\def \DOPHOT	{D{\sc o}PHOT}
\def \isis	{ISIS~2.0}
\def \hi   	{\ion{H}{1}}
\def\area	{${\rm deg}^2$}
\def\kpc	{\hbox{\rm kpc }}
\def\pc		{\hbox{ pc }}
\def\yr		{ \, {\rm yr}}
\def\peryr	{ \, {\rm yr^{-1} }}
\def\vlos	{ v_{\rm los} }
\def\lsim	{ \rlap{\lower .5ex \hbox{$\sim$} }{\raise .4ex \hbox{$<$} } }
\def\gsim	{ \rlap{\lower .5ex \hbox{$\sim$} }{\raise .4ex \hbox{$>$} } }
\def\solar	{ {\odot} }
\def\lsolar	{ {\rm L_{\odot}} }
\def\msolar	{ \rm {M_{\odot}} }
\def\rsolar	{ \rm {R_{\odot}} }
\def\surfmunit  { \rm {\, \msolar \, pc^{-2}} }
\def\HI		{{H{\sc I}}}
\def\mags	{{ \, \rm mag }}    
\def\abs	{ \hbox{ \vrule height .8em depth .4em width .6pt } \,} 
\def \tightenlines {\def\baselinestretch{1}\small}
\def\gtorder	{\mathrel{\raise.3ex\hbox{$>$}\mkern-14mu\lower0.6ex\hbox{$\sim$}}}
\def\ltorder	{\mathrel{\raise.3ex\hbox{$<$}\mkern-14mu\lower0.6ex\hbox{$\sim$}}}
\def\kpc{{\rm kpc}}
\def\day{{\rm day}}
\def\min{{\rm min}}
\def\lim{{\rm lim}}
\def\dls{{D_{\rm LS}}}
\def\dos{{D_{\rm S}}}
\def\dol{{D_{\rm L}}}
\def\te{{t_{\rm E}}}
\def\thetae{{\theta_{\rm E}}}
\def\re{{r_{\rm E}}}
\def\b{{\rm base}}
\def\ur{{u_{\rm r}}}
\def\tep{{t_{\rm E}^{\perp}}}
\def\bu{{\bf u}}
\def\cc{{\rm cc}}
\def\bmu{\hbox{$\mu\hskip-7.5pt\mu$}} 
\def\cm{{\rm cm}}
\def\rel{{\rm rel}}
\def\br{{\bf r}}
\def\bv{{\bf v}}
\def\bd{{\bf d}}
\def\e{{\rm E}}
\def\fb{F_{\rm B}}
\def\fs{F_{\rm S}}
\def\etas{\eta_{\rm \, see}}
\def\ce{{\rm ce}}
%
% -----------------------------------------------------------------------------
%  Title Page
%

\vskip 2cm

\title{Detection of Rotation in a Binary Microlens:\\ 
PLANET Photometry of MACHO~97-BLG-41
}

\vskip 0.5cm

\author{ 
M. D. Albrow\altaffilmark{1}, 
J.-P. Beaulieu\altaffilmark{2,3},
J. A. R. Caldwell\altaffilmark{4}, 
M. Dominik\altaffilmark{2}, 
B. S. Gaudi\altaffilmark{5},\\ 
A. Gould\altaffilmark{5}, 
J. Greenhill\altaffilmark{6}, 
K. Hill\altaffilmark{6},
S. Kane\altaffilmark{6,7}, 
R. Martin\altaffilmark{9},  
J. Menzies\altaffilmark{4}, 
R. M. Naber\altaffilmark{2},\\ 
K. R. Pollard\altaffilmark{1}, 
P. D. Sackett\altaffilmark{2,8}, 
K. C. Sahu\altaffilmark{7}, 
P. Vermaak\altaffilmark{4},  
R. Watson\altaffilmark{6}, 
A. Williams\altaffilmark{9}
}
\author{(The PLANET Collaboration)\\
~and\\
H.E. Bond\altaffilmark{7}, I.M. van Bemmel\altaffilmark{10,2}}

\altaffiltext{1}{Univ. of Canterbury, Dept. of Physics \& Astronomy, 
Private Bag 4800, Christchurch, New Zealand}
\altaffiltext{2}{Kapteyn Astronomical Institute, Postbus 800, 
9700 AV Groningen, The Netherlands}
\altaffiltext{3}{Institut d'Astrophysique de Paris, INSU CNRS, 98 bis 
Boulevard Arago, F-75014, Paris, France}
\altaffiltext{4}{South African Astronomical Observatory, P.O. Box 9, 
Observatory 7935, South Africa}
\altaffiltext{5}{Ohio State University, Department of Astronomy, Columbus, 
OH 43210, U.S.A.}
\altaffiltext{6}{Univ. of Tasmania, Physics Dept., G.P.O. 252C, 
Hobart, Tasmania~~7001, Australia}
\altaffiltext{7}{Space Telescope Science Institute, 3700 San Martin Drive, 
Baltimore, MD. 21218~~U.S.A.}
\altaffiltext{8}{Anglo-Australian Observatory, P.O. Box 296, Epping NSW 1710, Australia}
\altaffiltext{9}{Perth Observatory, Walnut Road, Bickley, Perth~~6076, Australia}
\altaffiltext{10}{European Southern Observatory, Karl-Schwarzschild-Strasse 2, 85748 Garching, Germany}

%----------------------------------------------------------------------

\begin{abstract}

We analyze PLANET collaboration data for \event, the only 
microlensing event observed to date in which the source transits 
two disjoint caustics.  The PLANET data, consisting of 
46 V-band and 325 I-band observations from five southern observatories,  
span a period from the initial alert until the end of the event.  
Our data are incompatible with a static binary lens, but are  
well fit by a rotating binary lens of mass ratio $q=0.34$ and 
angular separation $d\approx 0.5$ (in units of the Einstein ring radius) 
in which the binary separation changes in size 
by $\delta d = -0.070 \pm 0.009$ and in orientation by  
$\delta \theta = 5^\circ\hskip-2pt .61\pm 0^\circ\hskip-2pt .36$ 
during the 35.17 days between the separate caustic transits. 
We use this measurement combined with other observational constraints 
to derive the first kinematic estimate of the mass, distance,
and period of a binary microlens.  
The relative probability distributions for these parameters peak at a 
total lens mass $M\sim 0.3\,M_\odot$ (M-dwarf binary system), 
lens distance $D_{\rm L}\sim 5.5\,$kpc, and binary period $P\sim 1.5\,$yr.  
The robustness of our model is demonstrated by its
striking agreement with MACHO/GMAN data that 
cover several sharp features in the light curve not probed by the 
PLANET observations, and which did not enter our modeling 
procedure in any way.  Available data sets thus indicate that 
the light curve of \event\ can be modeled as a source 
crossing two caustics of a physically-realistic rotating binary so that,   
contrary to a recent suggestion, the additional effects of a 
postulated planetary companion to the binary lens are not required.

\end{abstract}


\vskip 2cm

\hangindent 2cm{\small 
$^*$Based on observations at:  Canopus Observatory, Tasmania, Australia; 
Perth Observatory, Bickley, Australia; 
the South African Astronomical Observatory, Sutherland, South Africa; 
the European Southern Observatory, La Silla, Chile; 
and the Cerro Tololo Inter-American Observatory, Cerro Tololo, Chile.}

%----------------------------------------------------------------------

\newpage
%---------------------------------------------------------------------------

\section{Introduction}\label{intro}

%---------------------------------------------------------------------------

One of the first microlensing events discovered was shown to be due 
to a binary lens (\cite{ogle7binary}), not surprisingly since 
most of the hundreds of events discovered toward the Galactic bulge  
are believed to be due to normal stellar lenses in the 
disk and bulge of the Milky Way 
(\cf, \cite{kirpac94}; \cite{zhao95}; \cite{alcock97}) 
and a sizable fraction of stars reside in binary systems.  
However, not all binary lenses can be recognized readily as such.   
For typical photometric sampling and precision, most light curves 
generated by double lenses will be indistinguishable  
from single-lens events if one lens component is sufficiently less massive 
than the other or if the components are separated by angular distances 
sufficiently smaller or larger than the Einstein ring radius.   
Nevertheless, several binary microlens events are recognized 
every year (\cf\ \cite{machobinaries99}), generally through    
sharp peaks in their light curves that betray the presence of  
extended caustic structure 
(closed and sometimes disjoint curves of formally infinite magnification 
on the sky) which are not generated by single point-lenses.  

Binary lenses have already demonstrated their importance to 
astronomy in several ways.   Timing the motion of a caustic across a  
finite background source of known size can be used to 
measure the relative proper motion of the lens and source. 
This was done by several groups who performed a joint analysis  
(\cite{jointMS98001} and references therein) of their data to demonstrate 
beyond reasonable doubt that the lensing system of event MACHO 98-SMC-1 
was not a constituent of Galactic dark matter, 
but a normal stellar binary in the Small Magellanic Cloud (SMC) itself.  
Caustic structure differentially magnifies background sources, providing 
a means of measuring limb-darkening coefficients for their stellar 
atmospheres, as has been done for a K-giant in the Galactic bulge 
(\cite{planetMB97028}) and an A-dwarf in the SMC (\cite{jointMS98001}).  
Mathematically, lensing planetary systems represent the 
extreme-mass-ratio subset of binary (or multiple) lenses 
(\cite{maopac91}; \cite{dominikextbin}).  
Two groups (PLANET, \cite{planetpilot}; MPS, \cite{mpsMS98001}) 
are conducting intense monitoring projects specifically designed to 
search for lensing planets. 
The efficiency with which planets can 
be detected must be determined separately for each light curve  
(\cite{gs2000}); massive planetary companions have been ruled 
out over a wide range of separations for two different microlenses 
(\cite{mpsmoaMB98035}; \cite{planetOB98014}).
 
The components of all physical binaries rotate about their 
common center of mass, 
a phenomenon that alters the shape and relative placement of 
caustic curves on the sky and thus, in principle, the form of any light curve 
generated by a binary microlens (\cite{dominikrotbin}).  
Nevertheless, rotation has not yet been demonstrated in any binary microlens.  
 
The unusual light curve of the Galactic bulge microlensing event 
MACHO 97-BLG-41 displays what appear to be source crossings of 
two disjoint caustic curves separated in time by about a month.  
The event has already been discussed by two groups with extensive 
photometric data sets: MACHO/GMAN, who report that they were unable 
to find a static or rotating binary model compatible with their data  (\cite{machobinaries99}), and MPS/Wise-GMAN (\cite{mpsMB9741}), 
who --- based on MPS+MACHO/GMAN data --- also discount rotation as an 
explanation for the features in the light curve, 
concluding that a three-body lens 
(binary + orbiting Jovian planet) is required to produce the 
light curve structure in \event.  
Theirs is the first claim in the literature for a planet orbiting 
a normal binary stellar system, as opposed to one component of 
a binary (\cite{butler97}; \cite{cochran97}) or a binary pulsar 
(\cite{PSRB1620}).  
Given the small chances of discovering such a system from microlensing
observations, Bennett \etal\ (1999) argued that they might well be common.

Here, we present and analyze the PLANET photometric data set for \event, 
which is completely independent of the MPS+MACHO/GMAN data set analyzed 
by Bennett \etal\ (1999).  We also find that no static binary model  
reproduces our data, but unlike Bennett \etal\ (1999) we conclude that 
a rotating binary model, with physically realistic parameters, does  
provide a satisfactory fit to all features in our light curve.
Since the two data sets are disjoint 
(and since the MACHO/GMAN data sample key
caustic-crossing regions that our data do not) we examine whether or not 
our solution can provide a satisfactory fit to their data.  
We find that our solution (derived without reference to any 
data but our own) is in reasonable agreement with MACHO/GMAN data, 
so that it appears likely that a single rotating-binary solution near 
the one we have found is consistent with all available data.
This result obviates the need for a third lensing 
component (planet) to explain the unusual light curve of \event.

We describe our data for \event\ 
and their reduction in \S2. Our general form for a rotating binary 
model is given in \S3.  Our modeling analysis is presented in \S4, 
including a demonstration of the non-viability of static binary models, 
and of the success of the rotating binary model 
in fitting not only our data but also the MACHO/GMAN data set for \event\ 
used in Bennett \etal\ (1999). 
In \S5 we present the physical and kinematical parameters for the binary 
lens derived from our analysis of our data, including the first statistical 
estimate for the mass and rotational period of a lensing binary made 
possible by our measurement of rotation in this system.
We conclude in \S6 that \event\ is the first microlensing event 
in which binary lens rotation has been demonstrated, 
and that it provides a credible, physically-realistic, and elegant  
model for this complex light curve.   
%---------------------------------------------------------------------------

\section{PLANET Photometric Data for MACHO-97-BLG-041}\label{data}

%---------------------------------------------------------------------------

An electronic alert of the microlensing event \event\  
was issued on 18 June 1997 by the MACHO\footnote{MACHO/GMAN Alert Homepage: 
http://darkstar.astro.washington.edu} team.  Many such alerts are made 
each year by the MACHO, 
OGLE\footnote{OGLE Alert Homepage: http://www.astrouw.edu.pl/$\sim$ogle/ogle2/ews/ews.html} 
and EROS\footnote{EROS Alert Homepage: http://www-dapnia.cea.fr/Spp/Experiences/EROS/alertes.html} collaborations, 
allowing a variety of follow-up observations to be undertaken by other teams. 
The \event\ event was described in the MACHO alert as in the direction 
of the Galactic bulge 
($\alpha = $17:56:20.7, $\delta = -$28:47:42) and possibly of short duration. 
The PLANET collaboration began monitoring immediately.
On 29 June 1997, MACHO reported that the light curve of the event
had flattened, but at a level above its baseline value, in a
manner consistent with expectations for a binary lens.
On 2 July 1997, PLANET\footnote{PLANET Collaboration Homepage:  
http://www.astro.rug.nl/$\sim$planet} reported an increase of 
brightness of a few percent, and on 23 July 1997 predicted 
that a caustic crossing was likely to occur the following day 
at JD - 2450000 = $654 \pm 1$; 
this was subsequently confirmed observationally by both MACHO and PLANET.

The PLANET data set for \event\ consists of photometric measurements 
in the $V$ band from three observing sites and in the $I$ band from five 
observing sites in the Southern Hemisphere during a four-month 
period directly after the initial alert.  
In addition, six measurements were taken after the event had 
reached baseline in April 1998 at the Canopus 1.0m in Tasmania 
with the same detector system as used in 1997.  
The data set thus comprises 8 separate light curves.  

After flat-fielding and bias subtraction, the source star was photometered 
using the image subtraction package \isis\ (\cite{isis2}).  In this 
procedure, pixel values are first interpolated to register all frames to an 
image chosen as the astrometric reference for that band and site. 
A photometric template is then formed for each band and site 
from several good-seeing, low-background frames.  
\isis\ uses a non-constant kernel with a fixed number of variable parameters 
to transform the template to each image by performing a least squares 
fit to derive the values of the kernel parameters.  
Profile-fitting photometry is then performed to measure the flux 
difference of variable objects between each frame and the template 
by convolving the point spread function (PSF) of the template with the best-fitting PSF matching kernel for each of the images. 
Frame-to-frame differences in exposure time, background, transparency, 
and the size and shape of the point spread function are automatically taken into account with this package, but since only flux {\it differences\/}  
are measured, the photometric system must be aligned with a 
standard system.   We have done this by performing a linear regression 
between ISIS flux differences and \dophot-reported fluxes (see below), 
and placing the latter on a standard system using standard stars. 
The \isis\ package produces a quality parameter to indicate the 
quality of the subtraction, which is a function of the signal-to-noise in the 
subtracted frame. To produce the cleaned data set used 
for analysis, we require that this quality flag is greater than 0.9, 
eliminating the poorest $\sim$35\% of the frames.  
In addition, we have discarded 9 other frames for insufficient exposure 
time, very strong disagreement with data taken 
at nearly the same time on a smoothly varying portion of the light curve, 
or sensitivity to $V$-band limb-darkening 
(3 $V$-band points too near a caustic), which we cannot constrain.  
The result is 325 $I$-band and 46 $V$-band measurements, for a grand 
total of 371 points in the high-quality, cleaned ISIS data set. 
A summary is given in Table~1.    

We have also reduced our data set with the PSF-profile-fitting 
package \dophot\ (\cite{dophot}), 
using fixed-position catalogs and the average of four or more stable, 
moderately bright and relatively uncrowded stars 
in the field as a relative flux standard (\cite{planetpilot}).   
Final cleaned \dophot-reduced data sets were formed by eliminating 
all trailed frames, frames that did not process properly, 
or frames whose image quality or signal-to-noise were not adequate to 
produce reliable (\ie\ \dophot\ type 11) measurements for the source star  
or the relative flux standard. 
Instrumental magnitudes of the SAAO~1m \dophot\ system 
were then calibrated against contemporaneous observations of Johnson-Cousins, 
UBV(RI)c E-region standards (\cite{menzies89}) at the beginning of the 
1999 observing season.  The instrumental magnitudes of other PLANET sites 
were brought to this system in the modeling process by introduction of 
an alignment parameter for every site and band.  This parameter is 
insensitive to widely different models and thus has no effect on the 
results we present here.   
Linear regression between the ISIS flux differences and 
the calibrated \dophot\ fluxes then allowed calibration of the ISIS dataset.  

\begin{figure}

% COLOR FIGURE AND CAPTION
\plotone{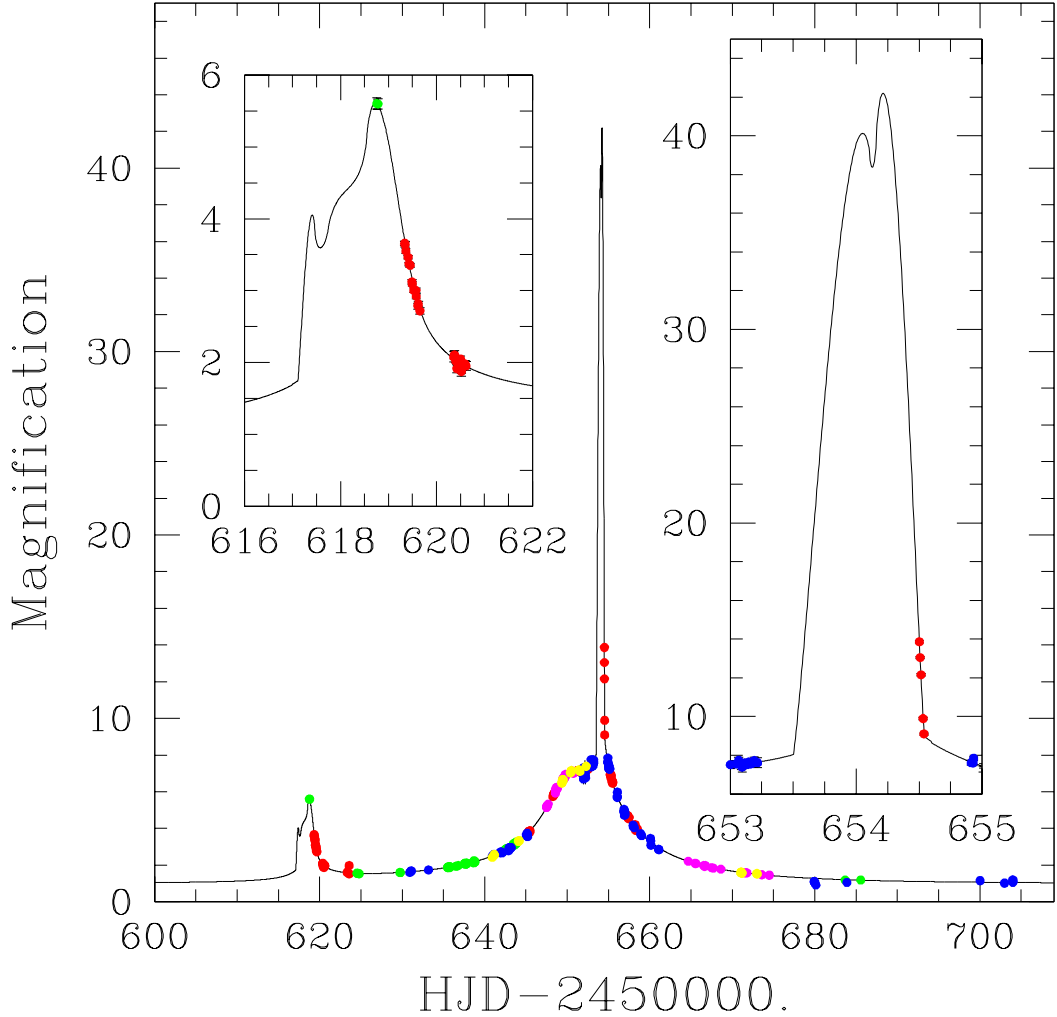}
\caption[]{\label{fig:lightcurves}
Our best rotating binary model shown with cleaned PLANET multi-site data 
in the $V$ and $I$ passbands 
({\it Green:~}Dutch/ESO 0.91m, LaSilla, Chile; 
{\it Red:~}SAAO 1m, Sutherland, South Africa;  
{\it Blue:~}Canopus 1m, Tasmania, Australia; 
{\it Yellow:~}CTIO 0.9m, Cerro Tololo, Chile;
{\it Magenta:~}Perth 0.6m, Bickley, Western Australia).   
Rescaled error bars (\S\ref{results}) are shown, 
but are generally smaller than the size of the points.
Insets are zooms on the two caustic crossing regions.
Plotted is magnification versus HJD$'$.
}

\end{figure}

The eight ISIS-reduced light curves of \event\ are shown in Figure~\ref{fig:lightcurves}, in which we have overplotted 
the best-fitting model presented in \S4.3 used to photometrically 
align the multi-site data.  
Due to differences in effective resolution at different sites and 
in different bands, each light curve can be subject to different 
amounts of non-lensed blended light.  
The model yields a best fit to this blended light 
so that we can plot the magnification of the 
lensed source light only in Figure~\ref{fig:lightcurves}.  
Plotting magnifications in Figure~\ref{fig:lightcurves} also allows 
presentation of $V$- and $I$-band data on the same plot, 
since the unlensed source magnification is by definition equal to unity 
irrespective of passband.\footnote{
Technically, the $V$- and $I$-band magnification curves can differ 
in regions such as caustic crossings where the passband-dependent 
source profile is important.  Since we have eliminated the few 
$V$-band points in these regions, this is not a concern for us here.} 
The time scale has been expressed in 
Heliocentric Julian Date (HJD) at the middle of the exposure, 
with a fixed offset: HJD$'$ = HJD$ - 2450000$. 
We use this timing system throughout. 

%---------------------------------------------------------------------------

\subsection{Reddening of the \event\ Field from the CMD}\label{cmd}

%---------------------------------------------------------------------------

\begin{figure}

% COLOR FIGURE AND CAPTION
\plotone{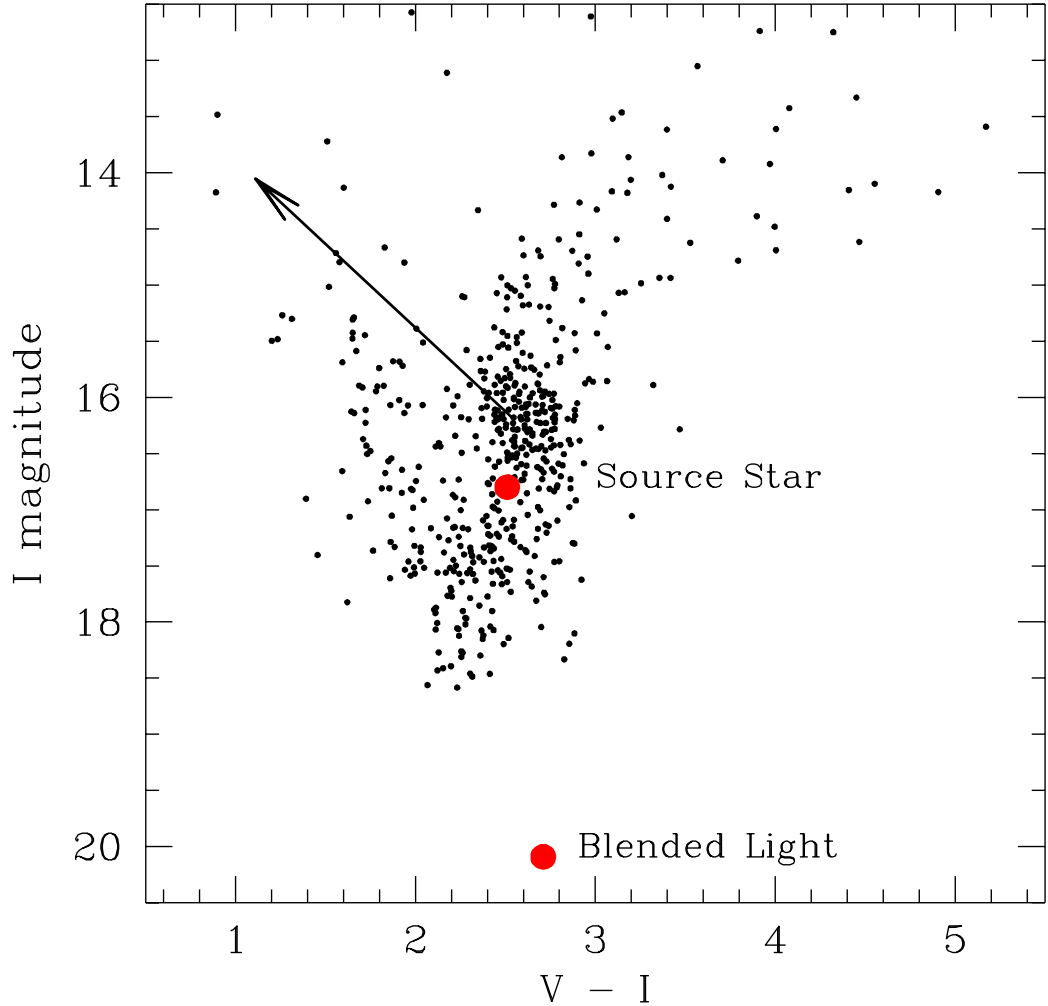}
\caption[]{\label{fig:cmd}
Calibrated color-magnitude diagram taken from SAAO~1m observations of a 
$1\arcmin \times 1\arcmin$ field centered on \event.  
The position of the unmagnified source and much fainter blend 
(as determined from our modeling) are indicated by the large solid dots.
The dereddening vector connecting the center of the observed red 
clump to the intrinsic clump characterized by 
Paczy\'nski \etal\ (1999) is also shown. 
}

\end{figure}

Figure~\ref{fig:cmd} displays a color-magnitude diagram (CMD) 
of a $1 \arcmin \times 1\arcmin$ field centered on \event\ derived from 
SAAO~1m images.   The position of the unmagnified source star 
is shown together with that of the blend 
(unresolved light along the line of sight), as determined from the 
models presented in \S\ref{rotbin}.  
Comparison of this CMD with one for a larger $5 \arcmin \times 5\arcmin$ 
field clearly indicates that the reddening is variable across the field. 

Using our standard-star calibration of this $1\arcmin$ subfield, 
we find that the center of the red clump is at $I_{\rm cl}=16.20\pm 0.05$
and $(V-I)_{\rm cl}=2.55\pm 0.05$.  Comparing this observed color to the 
dereddened color of $(V-I)_{\rm cl,0}=1.114$ derived by Paczy\'nski \etal\  
(1999), we obtain a value of the reddening $E(V-I)=1.44$.  
The reddening law $A_I=1.49 E(V-I)$ of Stanek (1996) then yields $A_I=2.15$.  
In \S\ref{params} we will use this value of reddening and extinction 
together with our best estimate of the color and magnitude of the source 
from our modeling to derive an angular size for the source and thence 
the relative lens-source proper motion. 

%---------------------------------------------------------------------------

\section{Model Parameters for a Rotating Binary Microlensing Event}\label{model}
 
%---------------------------------------------------------------------------

A static, binary lensing event of an unblended finite source can be 
described by the seven parameters:  
$d$, the angular separation between the two masses as a fraction of the
angular Einstein radius $\theta_{\rm E}$; 
$q \equiv M_2/M_1$, the mass ratio of the two components; 
the Einstein time $t_{\rm E} \equiv \theta_{\rm E}/\mu$, where $\mu$ is the 
relative proper motion between the source and the binary center of mass; 
$t_0$, the time of closest angular approach between the source trajectory and 
the binary center of mass; 
$u_0$, the angular separation between the source and binary center of 
mass at time $t_0$ in units of $\theta_{\rm E}$;
$\alpha$, the angle of the source trajectory on the sky relative to the 
binary axis; 
and $\rho_*$, the angular size of the source star in units of $\theta_{\rm E}$. 

Throughout, we use the Einstein ring radius of the total mass $M$, 

\begin{equation}
\theta_{\rm E}^2 \equiv {\frac{4 G M\dls}{c^2\dol\dos}} ~~~,
\label{eqn:thetae}
\end{equation}

\noindent
where $\dol$ is the observer-lens distance, $\dos$ is the observer-source 
distance, and $\dls \equiv \dos - \dol$. 
We center our coordinate system 
on the center of mass of the binary lens such that at time $t_0$ 
the lenses are aligned with the $x$-axis, with the more massive of the two, 
$M_1$ on the right.  The angle $\alpha$ is chosen between the positive $x$-axis 
and the source trajectory so that $M_1$ remains to the right of the source.

A binary system in which each partner travels on an 
ellipse about the common center of mass is described by the seven parameters 
of the reduced Kepler problem: the semi-major axis $a$, the period 
$P$, the eccentricity $e$, the phase $\phi$ (at time $t_0$, say), 
and the three angles of orientation.  
However, two of these, $a$ and one of the 
orientation angles, are already contained in the parameterization 
of a static lens microlensing event (in which distances are expressed in 
$\thetae$ and positions measured relative to the source trajectory), 
leaving five {\it additional\/} parameters specific to binary rotation.   
If the motion is circular ($e = 0$), 
three of these canonical parameters remain: 
$P$, $\phi$, and the inclination of the orbital plane with respect to the sky.  

We model our \event\ data with the simplest rotation model required  
to fit all the observed features, namely rectilinear relative motion 
of the lens components, which represents the next higher order term 
in lens motion compared to a static binary lens.  
Only the two components of this motion that lie in the sky plane 
are measurable since motion along the line-of-sight 
does not alter the caustic structure of the binary.  
We could expect to be able to detect the relative acceleration of 
the binary components only if we have a very strong detection of the 
rectilinear motion and the observations cover a sizable fraction 
of the rotation period.  We therefore model the binary rotation   
by adding only two degrees of freedom over that required for the static binary: 
$\Delta_x$ and $\Delta_y$, which measure the evolution of the separation 
vector during the course of the observations in the $x$- and $y$-directions. 
Under our assumption of rectilinear motion, the separation vector is thus 
 
\begin{equation}
\vec d = \vec d_0 + \vec\Delta \, \left( {\frac{t - t_0}{\te}} \right)
\label{eqn:deltas}
\end{equation}

\noindent 
where $\vec d_0 \equiv (d_0, 0)$ is the separation vector at time $t_0$, and 
$\vec \Delta \equiv (\Delta_x, \Delta_y)$.  
This formulation has the advantage that we need not assume anything 
about the eccentricity of the orbit.  The assumption of circular orbits would  
reduce the complexity of the full orbital solution, but is unlikely  
to be valid for the stellar binaries separated by several AU to which 
microlensing is sensitive (\cite{duqmaj1991}).  
Regardless of the eccentricity of the orbit, the first order term we 
are attempting to measure is the sky velocity of the separation vector. 
As we explain in \S\ref{rotbin}, however, 
it is both inconvenient and time-consuming to parameterize the search 
for rotating solutions in terms of the 9 (7 static + 2 rotating) 
canonical parameters.  
We will search for solutions using a different set of parameters, 
translating our final models back to the canonical system at the end 
of the process.

In addition to these parameters, for each of the $n=8$ light curves, 
model parameters must be assigned for the flux of the 
source star, $\fs$ and the unlensed background flux, $\fb$.  
These can vary among sites due to differences in calibration 
of instrumental magnitudes and effective spatial resolution. 
We also include a parameter $\etas$ 
for each light curve to fit a linear relation to the systematic trends  
between measured flux and seeing generally seen in our data 
(\cf\ \cite{planetOB98014}). 
The flux of light curve $i$ is thus modeled as 
\begin{equation}
F_i(t) = F_{{\rm S},i} \, A(t) + F_{{\rm B},i} + \eta_{{\rm \, see},i} \, 
[\theta_{{\rm FWHM},i}(t)-2'']~,
\label{eqn:fluxmodel}
\end{equation}

\noindent
where $\theta_{{\rm FWHM},i}$ is the FWHM of the seeing disk, and $A$ is the
magnification from the model.

Finally, since caustic structure can differentially magnify the 
finite source, we include a limb-darkening parameter $\Gamma_I$.  
Our limb-darkening parameter 
$\Gamma$ is related to the canonical linear limb-darkening coefficient $c$ 
through $c_I = 3 \Gamma_I / (\Gamma_I + 2)$ (Albrow \etal\ 1999b, Appendix B).
Because our $V$ data are not sufficient to constrain $\Gamma_V$, we 
eliminate this parameter from our modeling and the 3 $V$-band points 
that could be affected by it.

In sum, we require 9 physical parameters for the rotating binary, 
$3 n = 24$ for the multi-site photometric alignment, blending, and 
systematic seeing correlations, and 1 for source $I$-band limb darkening, 
for a grand total of 34 model parameters.  We describe how we search for 
solutions in this unwieldy parameter space in \S\ref{rotbin}.

%---------------------------------------------------------------------------

\section{Light Curve Analysis}\label{analysis} 

%---------------------------------------------------------------------------

The PLANET light curve of \event\ in Figure~\ref{fig:lightcurves} 
shows two temporally sharp features, one at HJD$' = $619 and 
a second at HJD$' = $654.  
Between these two broad features, the magnification is substantially 
less than $A = 3$, so that unless the source is strongly blended, 
it cannot be inside a caustic in this region (\cite{wittmao1995}). 

The leading rise and roll-over at HJD$' = $650 indicates an approach to one  
cusp (a sharp discontinuous feature in a caustic curve) 
before the actual crossing near a second cusp of the same caustic 
at HJD$' = $654.  
This crossing must have begun after HJD$'= 653.21694$, our last point on the
gentle leading shoulder of the second peak.  
The end of the crossing is marked by the ultra-steep 
decline in brightness near HJD$' = $654.5.
In the interval $654.50156\leq$HJD$'\leq 654.53788$, 
the magnification is falling quite rapidly at a linear rate of 
${{\rm d} A/ {\rm d} t} = -133.7 \pm 1.6\,\rm day^{-1}$.  

The first anomaly, which ends near HJD$'\sim 619$, initially triggered the microlensing alert of \event, 
then thought to be a normal short duration single-lens event.
The points in the interval $619.33864\leq$HJD$'\leq 619.65447$ fall 
smoothly with a slope of $d A/d t\sim -2.3\,\rm day^{-1}$ which, 
although rapid, is considerably slower than the rate seen during the 
caustic crossing near HJD$' = 654.5$.  
The concavity and moderate steepness of the light curve 
near HJD'$=619$ indicates that the source exits a caustic very 
close to or over a cusp, and approximately parallel to its axis.
The crossing must have begun before our first SAAO data point at 
HJD$'=619.34$.  
We note that the concavity of our data in this region 
is inconsistent with the particular three-body solution 
presented by Bennett \etal\ (1999), though these data alone do not 
preclude the possibility of a different three-body solution.

%---------------------------------------------------------------------------

\subsection{Failure of Static Models}\label{static}

%---------------------------------------------------------------------------

The number, shape, and relative orientation of the closed caustic curves 
generated by a binary lens are fixed by its mass ratio $q$ and separation $d$, 
so that information about the caustics from the light curve constrain 
these two quantities 
(Schneider \& Weiss 1986; Erdl \& Schneider 1993; Dominik 1999).  
For a given $q$, wide separations produce two diamond (4-cusped) caustics; 
close separations generate three caustics, one diamond and two triangular 
(3-cusped); intermediate separations produce one 6-cusped caustic.
In all cases, one of the caustics, called the central caustic, 
will be near the position of the more massive lens.  

Because the second peak in the light curve of \event\ sits  
atop a region of higher background magnification than the first,  
it is likely to be associated with the central caustic.   
The cusp approach at HJD$' =$650 {\it between\/} two widely separated 
crossings that are themselves near or over cusps  
does not support the notion that all three are due to the 
single caustic produced by an intermediate-separation binary.  
The proximity of the cusp-approach shoulder at HJD$' =$650 to the 
near-cusp crossing near HJD$' =$654 constrains the size of the 
central caustic to be rather small compared to the Einstein radius 
$\thetae$, an indication that the mass ratio $q$ is not near unity, 
$d \ll 1$, or $d \gg 1$.  
In the last category, caustics are symmetric about the line joining 
them (the binary axis).  Since the light curve of \event\ is strongly 
asymmetric about the second caustic crossing, the source trajectory must 
be poorly aligned with the binary axis and thus could not cross 
the other caustic of a wide binary unless the binary were rapidly rotating.
On the other hand, the triangular caustics of close binaries with mildly 
unequal masses are naturally positioned at an angle with respect 
to the axis of symmetry of the central caustic. 
We therefore begin by exploring binary models of this type, asking whether 
the light curve could be produced by the source traversing the 
magnification pattern of a close binary whose caustics 
remain motionless during the course of the event.

\begin{figure}

% COLOR FIGURE AND CAPTION
\plotone{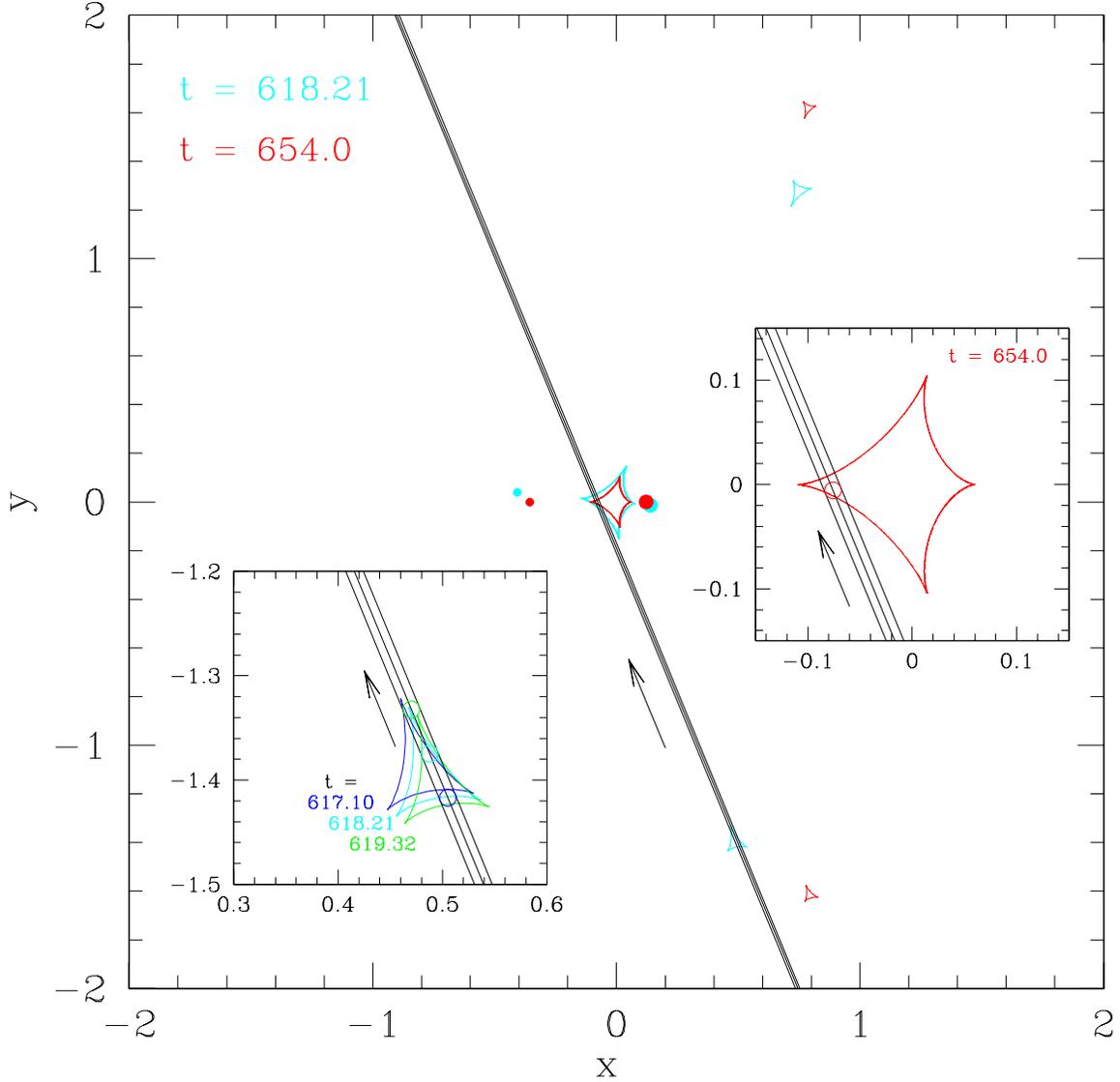}
\hglue 4cm
\caption[]{\label{fig:caustictopo}
The caustic topology of our best-fitting rotating model is shown 
at times near the first and second caustic crossings.  
Units are angular Einstein radii $\thetae$. 
The straight line shows the source trajectory and the arrow its  
direction of motion.  The binary components are shown as small dots.  
As time progresses, the binary rotates counterclockwise 
(causing the caustic pattern to do the same) 
and the component lenses move closer together (increasing the separation 
between the triangular outer caustics and the central caustic). 
The zooms better illustrate the source (circle) trajectory 
as it crosses the central caustic (right) and triangular caustic (left).  
The triangular caustic movement during the entire first crossing 
is shown.  The source diameter is indicated by the distance between 
the parallel lines.  
}

\end{figure}

Strikingly, and despite the fact that our data do not cover either
caustic crossing very well, we find that no static binary provides a
good fit to the data.  Static models that reproduce the structure of 
the strongly asymmetric second peak cannot reproduce the first 
caustic region because the source fails to pass through either 
of the smaller triangular caustics at an earlier time.  
(The topology of the central and triangular caustics can be appreciated 
by examining Fig.~\ref{fig:caustictopo}.) 
For the same reason, static models that fit 
the first light curve feature are very poor fits to the second. 
A static binary microlens model simply cannot explain the features of our 
data for \event.  Bennett et al.\ (1999) came to the same conclusion 
based on their data for this event, which also led them to search for a 
rotating binary solution.  Finding no viable rotating model, 
they concluded that the first caustic crossing was not due 
to one of the binary's own ``natural'' caustics, 
but to additional caustic structure induced by a planet.  
We turn now to the viability of a rotating binary model for \event.

%---------------------------------------------------------------------------

\subsection{Success of Rotating Models}\label{rotbin}

%---------------------------------------------------------------------------

For certain orientations and data samplings, planetary anomalies 
may be confused with those caused by triangular caustics of close binaries 
with higher mass ratio.  
Although the position of the triangular caustic in  
static models of \event\ is too misaligned with the source trajectory 
to explain the first caustic peak, 
its {\it size\/} is compatible with the duration of the first peak, 
in both our data and those used by Bennett \etal\ (1999).  
We found that models that simply rotate  
this caustic pattern on the sky yield many of 
the essential features of the light curve.  
Encouraged by this, and recognizing that caustic crossings widely 
separated in time should offer the best chance for detecting 
binary rotation, we proceeded to a quantitative search for rotating 
binary models as described in the following section. 

%---------------------------------------------------------------------------

\subsubsection{Search Strategy}\label{strategy}

%---------------------------------------------------------------------------

As we noted in \S\ref{model}, a rotating binary observed for a 
small fraction of its period can be approximated as two masses moving 
in straight lines about their common center of mass.  
We parameterize this system by beginning with the 7 parameters 
that characterize a static binary event ($d_0,q,t_\e,t_0,u_0,\alpha,\rho_*$), 
and add $\Delta_x$ and $\Delta_y$ to account 
for the two dimensions of (projected) relative velocity of the binary.
Once a model characterized by these 9 parameters is
chosen, the best fit for the additional photometric parameters 
$F_{{\rm S},i}$, $F_{{\rm B},i}$ and $\eta_{{\rm \, see},i}$ can be determined
by linear fitting, which requires a negligible amount of computer time.
Hence, the complexity of the problem is basically determined by the 
character of the 9-parameter space.

In practice, it is quite difficult to search this canonical parameter space by 
means of standard $\chi^2$ minimization routines, in part because
the dimensionality is large, but mainly because the $\chi^2$ surface 
is highly irregular.
A useful approach is to seek an alternative empirical 
parameterization of the problem
in which some of the parameters are fixed, or at least highly constrained
by subsets of the data.  For example,  Albrow et al.\ (1999b) showed 
that for static binaries with a well-sampled fold-caustic crossing, 
it is possible to reduce the primary search space from 7 to 5 dimensions 
by choosing the parameterization carefully.  
Moreover, they were able to restrict the 5-dimensional
search to points outside of caustics (for which finite source effects 
can be ignored or handled analytically so that the magnification can
be evaluated 100 or even 1000 times faster than for caustic-crossing points),
thereby rendering a 5-dimensional brute-force search tractable.

Here we develop a similar strategy.  
Three quantities are relatively well determined by 
cursory inspection of our data: the time the center of the source  
exits the first caustic $t_{\cc,1}$, the time the trailing limb exits 
the second caustic $t_{\ce,2}$, and the minimum separation from the 
source center relative to the first cusp $u_{\rm c,1}$.  
We construct a parameterization with these three quantities fixed 
at initial guesses derived from inspection of the data 
so that the search can be restricted to the remaining six parameters.  
More specifically,  by linearly extending the light curves from 
just before and just after the end of the crossing 
(see e.g.\ Afonso et al.\ 1998), we find that the end of the second
crossing is at $t_{\ce,2} = t_{\cc,2}+\Delta t_2=654.544\pm 0.001$ days.
Here $\Delta t_2$ is the radius crossing time of the source 
over the second caustic so that $t_{\cc,2}$ is the time the source 
center egresses.  

We begin our fitting process by 
assuming that the center of the source passes exactly over the
first cusp at $t_{\cc,1}=619.34$, our first data point in the series 
of SAAO 1m points on the concave decline. After one iteration, 
we find that a better estimate is that the center passes closest to the 
cusp of the first caustic at $t_{\cc,1}=619.15$.
The time between caustic crossings is then 
$\delta t\equiv t_{\cc,2}-t_{\cc,1} \simeq 35.18$ days, 
where we have made the evaluation based on our final measurement of 
$\Delta t_2$.
The six remaining parameters allowed to vary in our search 
are $(d_{\rm mid},q,\ell_2,\Delta t_2,\delta d,\delta\theta)$.
Here $d_{\rm mid}$ is the average of the separations $d_1$ and $d_2$ 
at the times of the caustic crossings $t_{\cc,1}$ and $t_{\cc,2}$,
$\ell_2$ is the coordinate of the egress point on the central diamond   
caustic relative to the cusp (as measured along the cusp bisector)  
in units of $\thetae$, $\delta d$ is the change in the binary 
separation $d$ during $\delta t$, and $\delta \theta$ is the change 
in binary orientation during $\delta t$. 
In our sign convention, $\delta d > 0$ and $\delta \theta > 0$ 
imply a binary lens that is separating and rotating  
counterclockwise on the sky with time.  

We initiate our search for rotating solutions by considering only data
from $t>621$ (and encapsulate all the information from the earlier data
in the assumption that the source crosses the cusp at $t=t_{\cc,1}$).
All but 5 of these points are outside the caustics and so require only
a single call to the lens solver to evaluate the magnification.  
Hence, the entire light curve can be evaluated quickly ($\sim$1~second 
on our system), permitting a very rapid search of the 6 dimensional space.
For each such model near a minimum in $\chi^2$, we examine a grid of
models for the early data points $t<621$.  This grid is characterized by
two parameters, namely the two parameters we had held fixed earlier: 
the time of the first caustic exit and the position of the source relative 
to the cusp at this crossing.  A full integration over the source 
is required for most of these early points, but since the search is on 
a 2-dimensional grid, it can be performed in about 4 minutes.  
For each 6-dimensional model, we add the additional $\chi^2$ from the best 2-dimensional grid to obtain the overall $\chi^2$.

%---------------------------------------------------------------------------

\subsubsection{Results}\label{results}

%---------------------------------------------------------------------------

Joint fits to the multi-site multi-band cleaned ISIS data were performed 
using the empirical parameters and discrete-grid strategy 
described in the previous section.  
The parameters of this minimum were then converted to the canonical 
parameterization, and the minimum checked  via a more conventional, 
but time-consuming, search using a downhill simplex method 
beginning with a seed near the empirical solution.  
Since the formal errors reported by photometric reduction packages 
often do not represent the true scatter in crowded-field data, 
systematic effects are often present that depend on site and target star. 
We therefore multiply the error bars of each of our 8 light curves by 
scaling factors $\sigma/\sigma_{\rm ISIS}$ that force $\chi_i^2/(N_i - 3) = 1$, 
where $N_i$ is the number of data points in light curve $i$.   
(Three degrees-of-freedom are already used by the 
$F_{{\rm S},i}$, $F_{{\rm B},i}$ and 
$\eta_{{\rm \, see},i}$ fitting parameters.)   
These error bar scaling factors, given in Table~1, are necessary to ensure 
that light curves with substantially underestimated photometric 
uncertainties are not given more weight than they are due in the joint fit.  
As an aside, we note that these scaling factors are closer to 
unity for ISIS reductions than for \dophot\ reductions 
in most cases we have tested, apparently indicating that 
ISIS-reported errors more accurately reflect the true scatter in 
the photometric data.  

Photometric parameters that depend on site, detector, and passband 
are presented in Table~1 for the resulting best-fitting model to 
the cleaned ISIS data set.  
In Table~2 we list the nine empirical geometric fitting 
parameters and their corresponding canonical counterparts for both 
data sets.  Uncertainties are presented for the fitted parameters; 
canonical parameters are expressed to more digits than are significant 
since this precision is required to reproduce the solution described 
by the fitted parameter set.    

To specify a solution to a multi-parameter problem to better than
$\Delta\chi^2=1$, one must give each parameter $i$ to a precision 
$B_{ii}^{-1/2}$, where $B\equiv C^{-1}$ and $C_{ij}$ is the covariance
matrix.  If the parameters are highly correlated, as they are for the
canonical parameters of a rotating binary, then
$B_{ii}^{-1/2}\ll C_{ii}^{1/2}$, so that the individual parameters must be 
specified to much higher precision than the statistical errors.  When  
we convert from the empirical to canonical parameters, we find that numerical
errors enter that are small compared to $C_{ii}^{1/2}$ but  comparable
to $B_{ii}^{-1/2}$.  We therefore minimize $\chi^2$ for the canonical
parameters using downhill simplex with the transformed solution as our
starting point.  This refined minimum is given as the last column of Table 2.  

Most of the parameters in the two solutions are very similar, 
but $c_I$ differs by $\sim 0.6\sigma$.  Thus our two 
minimization routines associated with our two parameterizations agree
at $\Delta\chi^2<1$ but not at $\Delta\chi^2\ll 1$, indicating that one
or both has numerical errors at this level.  Since the numerical errors
are smaller than the statistical errors, they are of no concern here, 
but would be if the quality of the data could be improved by a factor ten.

We have computed the full matrix of correlation coefficients for our
empirical, geometric fitting parameters and have found that the    
parameters $Delta t_2$ and $\Gamma_I$ are anti-correlated at the 
94\% level.  This strong anti-correlation arises
because the majority of the constraint on $\Gamma_I$ comes from the 5 SAAO
data points near HJD$'=654.5$, for which the limb of the source was exiting
the second caustic.  Measurements near the end of a caustic crossing
are subject to a degeneracy between the source size and limb darkening,
such that a large source size can be partially compensated for by a smaller
amount of limb-darkening, and vice versa.  This may restrict our ability 
to predict the light curve shape while the source is inside the caustic 
region, though, as we show in \S\ref{machogman}, this restriction is 
not severe.

The model light curve produced by our best model is shown in 
Figure~\ref{fig:lightcurves} with the PLANET data superposed.  
As the expanded views on the caustic regions show, 
this rotating binary fit is excellent; 
the solution reproduces our data in every detail.  
Binary lens rotation in \event\ is not 
only detected, but rather well determined by the PLANET data: 
$\delta d= -0.070 \pm 0.009$ and 
$\delta\theta = 5^\circ\hskip-2pt .61\pm 0^\circ\hskip-2pt .36$ represents  
$\sim$13\% and $\sim$6\% precision (1$\sigma$) for the magnitude of the 
radial and azimuthal motions on the sky, respectively. 
We find no other rotating solutions with similar $\chi^2$, and 
suspect that none exists since the covariance matrix for the empirical 
parameters varies smoothly over a very wide range in $\chi^2$ 
during our search.
 
\begin{figure}

% COLOR FIGURE AND CAPTION 
\plotone{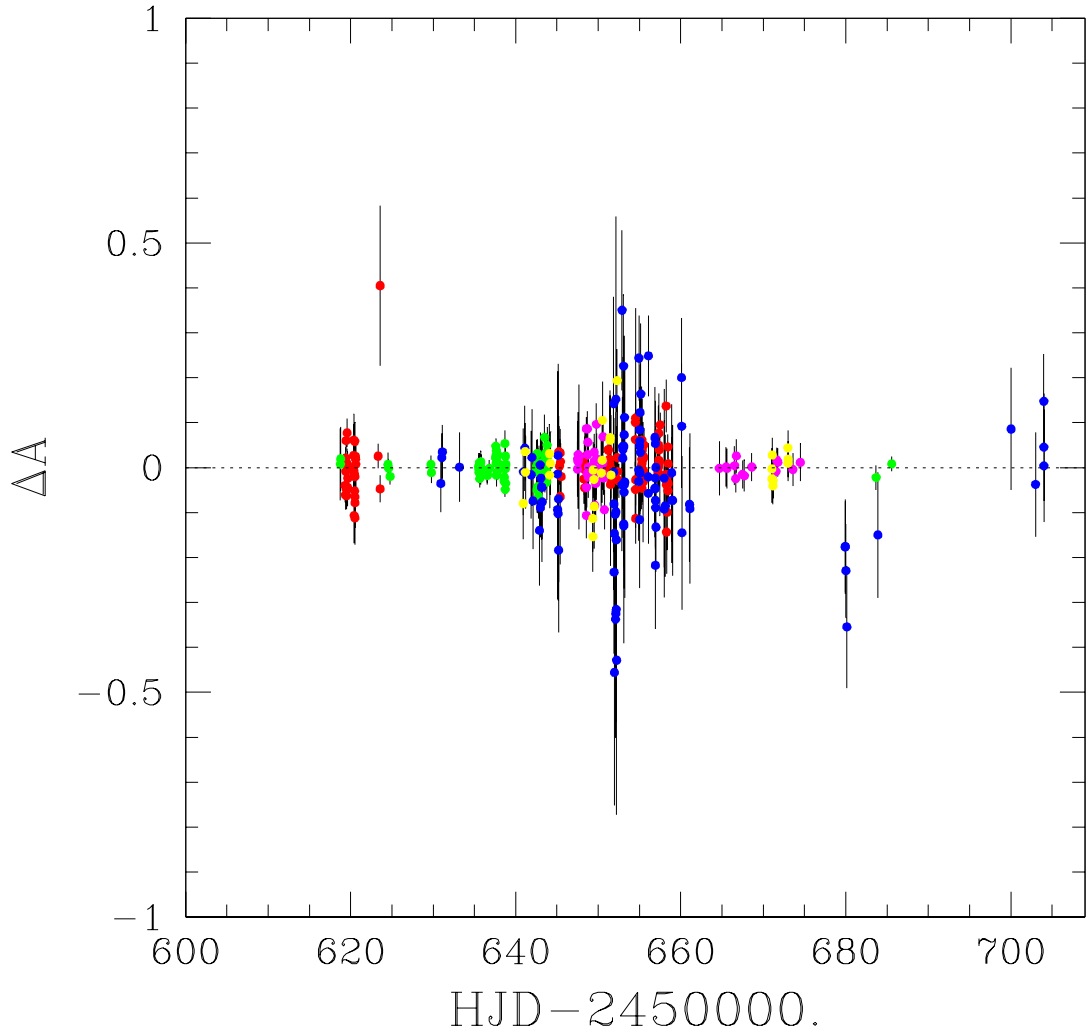}
\caption[]{\label{fig:residuals}
Magnification residuals (data $-$ model) of the cleaned PLANET data set from 
our best-fitting rotating binary model are shown as a function of HJD$'$. 
The color coding follows that of Fig.~\ref{fig:lightcurves}. Error bars have 
been rescaled to account for systematics (\S\ref{results}). 
Full moon occurred at HJD$' = 620$ and HJD$' = 649$.
}

\end{figure}

The residuals of the ISIS cleaned data set from our best model 
are shown in Figure~\ref{fig:residuals}.  
Since we have assumed rectilinear motion of the lens components, 
systematic residuals with time measured from 
$t = (t_{\cc,2}+t_{\cc,1})/2$ might indicate accelerated motion.   
The absence of systematic effects of any kind in the residuals 
indicates that our assumption of rectilinear motion is 
a good approximation, but also implies that our data for \event\ 
are not sufficient to detect the angular acceleration of the 
rotating binary components. 
  
Our best fits indicate that the unblended color and magnitude
of the source are $(V-I)_{\rm S}=2.51 \pm 0.03$ and $I_{\rm S}=16.82 \pm 0.02$, respectively. These uncertainties are independent because the first 
depends only on the ratio of $V$ and $I$ flux, while the latter 
depends on other model parameters.  
The blended light, whether from unrelated stars 
along the line of sight or from the lens itself, is quite faint: 
our best model yields blend magnitudes of $V_{\rm B}=22.8$, with errors 
large enough to make the detection unreliable,  
and $I_{\rm B}=20.09 \pm 0.47$.  
The small amount of blended light places a constraint on the mass of 
a stellar lens: a primary more massive than $M_1 = 1.4\,M_\odot$ 
(and hence a binary more massive than $M = 1.9\,M_\odot$) at any 
position between us and the Galactic bulge would produce a brighter blend 
than is observed.  A spectrum taken by Lennon \etal\ (1997) 
during the second crossing may be able to exclude smaller binary masses 
if the blue ($< 4500$\AA) unpublished portion of the spectrum fails 
to indicate the flux expected from less massive (hotter) stars. 

Lennon \etal\ (1997) deduce from their high signal-to-noise spectrum  
that the source star is a cool giant with log~$g = 3.2 \pm 0.3$, 
$T_{\rm eff} = 5000 \pm 200$, and [Fe/H]$ = -0.2 \pm 0.2$~dex.  Interpolating 
between the model limb-darkening grids of Claret, D\'iaz-Cordov\'es, and 
Gim\'enez (1995), we estimate that such a source star would be 
expected to have a linear limb-darkening coefficient of $c_I = 0.56$.
Limb darkening of the source, which affects the SAAO 1m data in the 
sharp decline after the second caustic crossing at HJD$'=654.5$ and 
(to a lesser extent) the LaSilla data after the first crossing at 
HJD$'=618.8$, can be measured from our light curve data alone.   
We find $\Gamma_I = 0.42 \pm 0.09$, corresponding to a linear limb-darkening 
coefficient of $c_I = 0.52 \pm 0.10$, in agreement with atmospheric models 
for a source of the type deduced by Lennon \etal\ (1997).  
Our $V$ data do not meaningfully constrain $\Gamma_V$. 
The dereddened source color of $(V-I)_{\rm S,0}=1.07 \pm 0.05$ 
we derive in \S\ref{params} together with isochrones of 
Bertelli \etal\ (1994) suggests $T_{\rm eff} = 4750 \pm 150$, 
somewhat cooler but consistent with that of Lennon \etal\ (1997).  

%---------------------------------------------------------------------------

\subsection{Comparison with MACHO/GMAN Data}\label{machogman}

%---------------------------------------------------------------------------

Our rotating binary solution makes use of only PLANET data, which
unfortunately sample only two neighboring points near the first caustic
and five points at the very end of the second caustic.  Nevertheless,
since our solution predicts the full light curve over these caustic 
intervals, we compare our prediction with the MACHO/GMAN data for this 
event, kindly provided to us by A. Becker 
(1999, private communication).  These data were presented in 
Alcock \etal\ (1999) and incorporated into the
analysis of Bennett \etal\ (1999), although with two different treatments  
of the systematic uncertainties.  

\begin{figure}

% COLOR FIGURE AND CAPTION 
\plotone{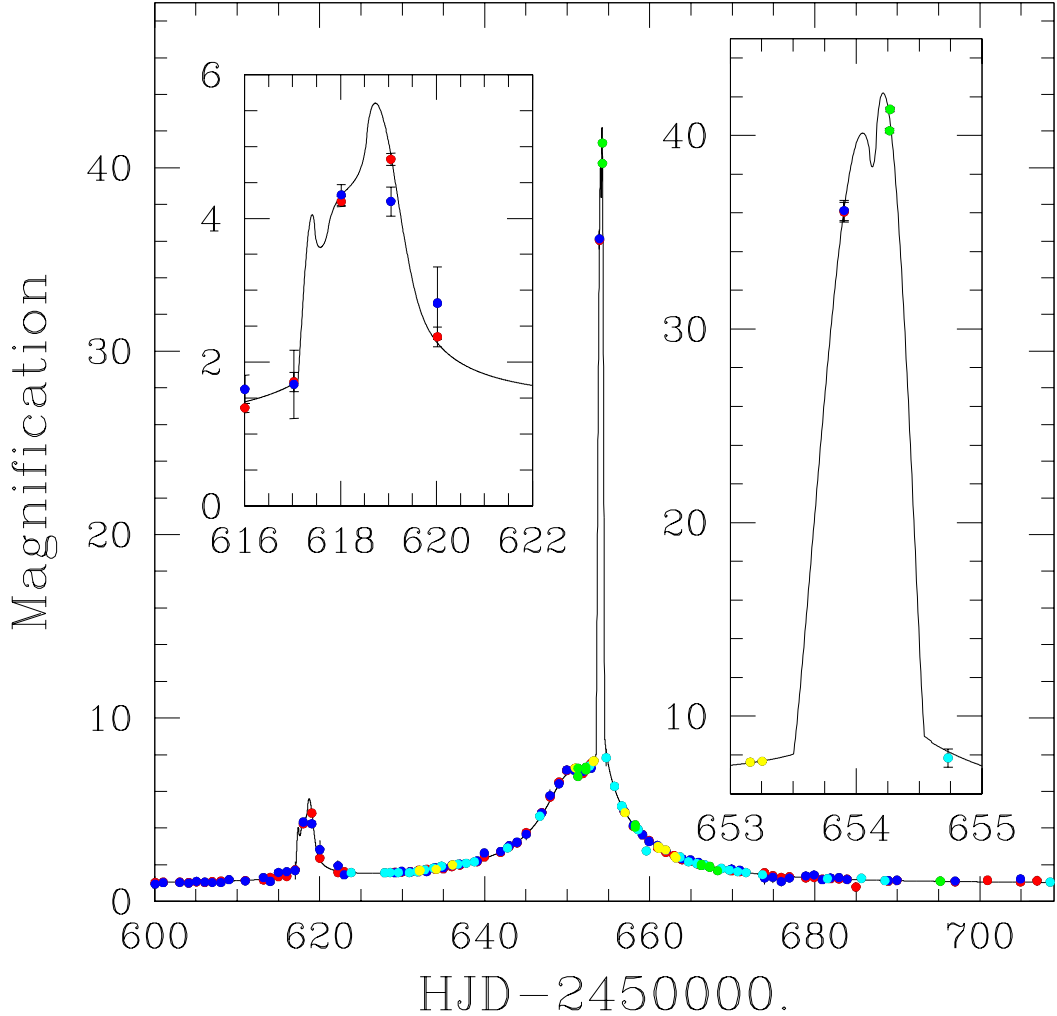}
\caption[]{\label{fig:machogmandata}
The curves show the magnification of MACHO 97-BLG-41 as a function of time
derived from the best-fitting model to the PLANET data, and are thus  
identical to those in Fig.~\ref{fig:lightcurves}.  The points are data from
the MACHO/GMAN collaboration kindly provided by A.\ Becker (1999, private
communication).  These data points did not enter the fit in any way: they
are simply superposed on the model curve after
undergoing a linear transformation to convert from flux to magnification. 
Color coding indicates observing site and passband 
({\it Red:~}MACHO~R; 
{\it Blue:~}MACHO~B;   
{\it Cyan:~}CTIO~0.9m~R; 
{\it Yellow:~}MSO~0.8m~R;
{\it Green:~}Wise~1m~R).   
Insets show the same two zooms displayed in Fig.~\ref{fig:lightcurves}. 
The agreement in regions devoid of PLANET data, 
near the beginning of the first caustic crossing and 
throughout the bulk of the second crossing, is particularly striking.
}

\end{figure}

\begin{figure}

% COLOR FIGURE AND CAPTION 
\plotone{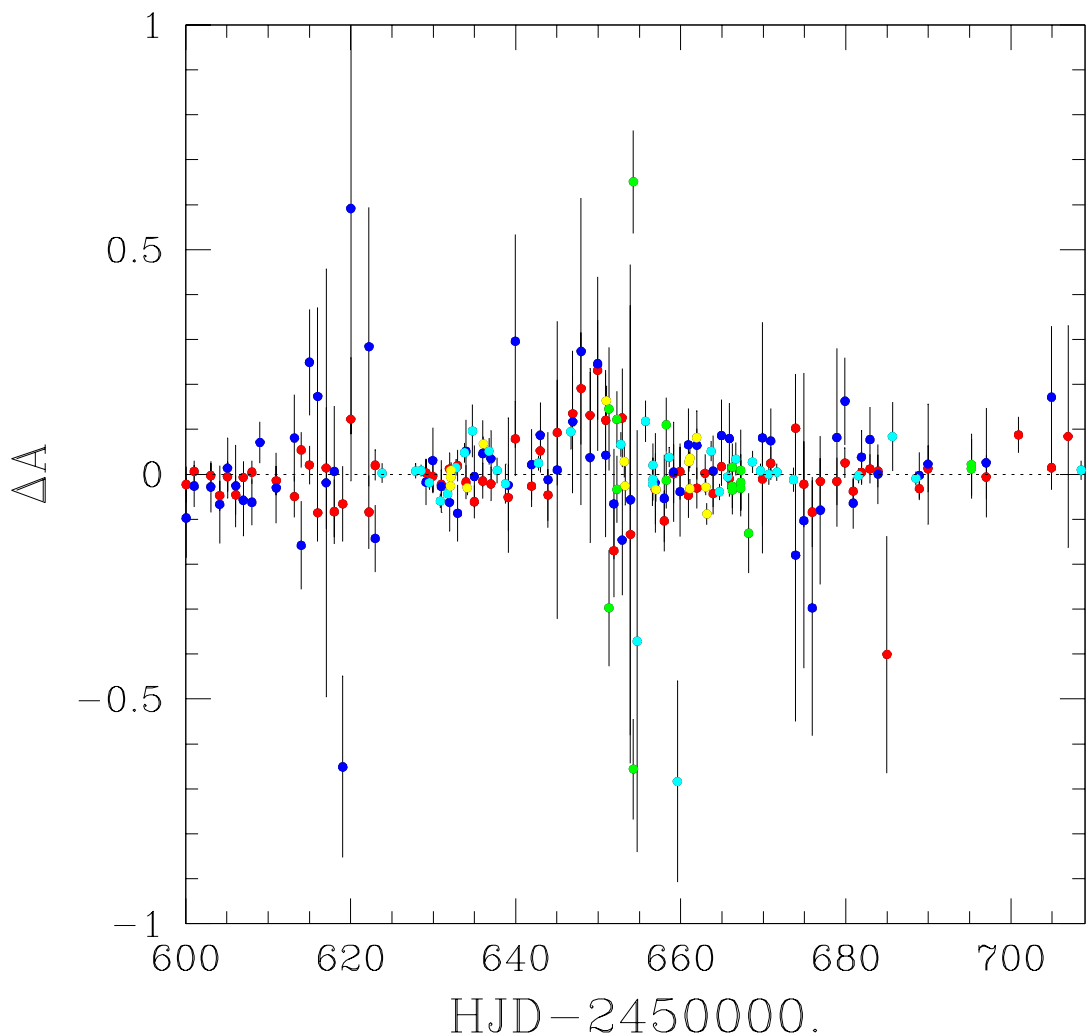}
\caption[]{\label{fig:machogmanresiduals}
Magnification residuals of the MACHO/GMAN data from our best-fitting model 
to the PLANET data only are shown as a function of HJD$'$. 
The color coding follows that of Fig.~\ref{fig:machogmandata}.
MACHO error bars have been multiplied by a customary 1.4 to account 
for systematics.  Error bars for GMAN data are those reported by the 
crowded field reduction packages;  
no attempt has been made here to increase the size of the GMAN error 
bars to account for systematic uncertainties (\S\ref{machogman}).  
Full moon occurred at HJD$' = 620$ and HJD$' = 649$.
}

\end{figure}

We show the comparison between the PLANET model and MACHO/GMAN data 
for the full light curve of the event and for zooms of the first
and second caustics in Figure~\ref{fig:machogmandata}.
In order to determine the magnifications from the flux levels
provided to us, we fit the fluxes to a function
$F(t)= A(t) \fs\ + \fb$ where $\fb$ is constrained to be non-negative 
and $A(t)$ is the
magnification in our model at the time of observation $t$.  
The parameters $\fs$ and $\fb$ are the minimum required to convert 
from observed flux to magnification.  No other fitting parameter 
was altered to accommodate the MACHO/GMAN data.  
The light curve is calculated using the $I$ band limb-darkening
coefficient $\Gamma_I=0.42$. Although the MACHO/GMAN data were taken  
in other bands, this affects only a few points and those by only 
a small amount.  The residuals of the MACHO/GMAN data from our 
model are shown in Figure~\ref{fig:machogmanresiduals}.

Without accounting for possible systematics in the data 
through error bar rescaling, we find that our model produces a 
$\chi^2 = 2787$ for 1557 MACHO/GMAN data points.  The same model produces 
a $\chi^2 = 2369$ for the 1317 MACHO/GMAN data points before or after
the microlensing event, indicating that the large $\chi^2$ is likely due 
to underestimated error bars and not due to problems in the model.  

Overall, our model prediction derived from 
PLANET data is in good agreement with the MACHO/GMAN points, even though 
their data did not enter the fitting process in any way, 
including in the choice of initial model parameters.  
Even portions of the MACHO/GMAN light curve that are devoid of 
PLANET data are generally well matched by the PLANET model.
The one exception is the {\it slope} of the two Wise~1m points at
HJD$'=654.26$, which are separated by 0.004 days.  While the model 
reproduces the midpoint of these two points almost exactly, it
predicts a change of $-$0.5\%, whereas the measured flux changes
by $2.7\pm 0.4\%$, a formal discrepancy of $8\,\sigma$, which is 
discernible but not obvious in Figure~\ref{fig:machogmandata} due
to horizontal compression.  This conflict cannot be resolved
by minor adjustments to the model: matching the data would require
putting these two points in a very different place relative to
the caustic structure, which cannot be accomplished without
significant changes in the model parameters away from the
global minimum found using the PLANET data.  Hence,
if the PLANET model is even approximately correct, the
errors on these two Wise points must be severely underestimated.

The MACHO error bars that we have used and shown 
in Figures~\ref{fig:machogmandata} and \ref{fig:machogmanresiduals} 
are those reported by the SoDoPhot photometry program with an 
additional 1.4\% added in quadrature; this is the standard procedure 
used by the MACHO team to account for systematic uncertainties. 
The uncertainties displayed for the GMAN data from CTIO and Wise  
observatories are those reported from the ALLFRAME photometric package 
used for their reduction, and have not been modified in any way. 
As we discuss in \S\ref{data}, when fitting models to data 
with mis-estimated errors, rescaling must be done  
in order to weight one data set properly against another 
to obtain valid statistical error estimates for the 
derived parameters.  A.\ Becker (1999, private communication) reports 
that in the MACHO/GMAN analysis (Alcock et al.\ 1999) the GMAN error bars 
were multiplied by a factor of 1.5, while D.\ Bennett 
(1999, private communication) reports that the MPS analysis 
(Bennett et al.\ 1999) added 1\% in quadrature to the uncertainties 
for GMAN data sets.  
Since we do not have access
to the raw data or even to their global characteristics (seeing,
background, exposure time), we cannot offer an independent judgment.  
However, because we are not fitting models to 
these data, but only reporting their residuals from our model fit to
PLANET data, there is no compelling reason to rescale.  
For completeness, we note that the residuals have 
$\chi^2=2787$ for 1557 data points with MACHO rescaling only (shown), and 
$\chi^2=2645$ and $\chi^2=2622$ under the two GMAN rescaling schemes 
listed above.  Since 1489 of the total 1557 points are from the MACHO 
team, the difference in GMAN data rescaling has a small effect on the total 
$\chi^2$.

One might ask how the PLANET model is able to fit MACHO/GMAN data in 
regions interior to the caustic where we have no data and where knowledge 
of the source size (relative to the Einstein ring) is crucial.  
The excellent PLANET coverage over the larger anomaly allows determination 
of $q$, $d_{\rm mid}$, and the trajectory over the second caustic quite 
accurately, so the path of the center of the source through the 
caustic region is well known.  Our data at the end of the 
second caustic crossing fix the time at which the limb of the source 
exits the caustic.  Together, the timed trajectories of the source center 
and its limb allow us to determine the source size with a precision 
sufficient to match MACHO/GMAN caustic data.

%---------------------------------------------------------------------------

\section{Physical and Kinematical Parameters of the Lens System}\label{params}

%---------------------------------------------------------------------------

The proper motion is given by $\mu = \theta_*/t_*$ where $\theta_*$ 
is the angular radius of the source and $t_*$ is the time required for the 
source to travel its own radius.  Since the source trajectory and 
total time for the source to cross the caustic are well-determined by 
our data, $t_*$ is known accurately from our model.   
We find $t_* = \Delta t_2 \sin{\phi} = (0.193 \pm 0.005)$~days, 
where $\Delta t_2$ is the time for the egressing source radius to cross the 
second caustic and $\phi = 88.2$\deg\ is the angle between 
the source trajectory and the caustic at this point.

From our best-fitting solution, we also find the unblended color and magnitude
of the source, $(V-I)_{\rm S}=2.51$ and $I_{\rm S}=16.82$, which imply 
intrinsic colors of $(V-I)_{\rm S,0}=1.07$ and $I_{\rm S,0}=14.67$ 
using the extinction $A_I = 2.15$ and reddening $E(V-I) = 1.44$ 
we derived in \S\ref{cmd} for this field.  
In order to determine the angular radius of the source, we
rederive the empirical surface brightness-color relation of van~Belle (1999),
using the same data set but restricted to the domain $2.0\leq V-K\leq 3.5$.
These data were generously provided to us by G.~van~Belle (1999, 
private communication) in advance of publication.  We obtain
\begin{equation}
\log(2\theta_*) + V/5 = 1.2885 \pm 0.0063 + (0.2226 \pm 0.0133)[(V-K)_0-2.823]~,
\label{eqn:newbelle} 
\end{equation}
where the uncertainties are expressed so that they are uncorrelated, and 
the half-interval of the uncertainty corresponds to $\Delta\chi^2=1$.  
Note that throughout we have used $\theta_*$ to denote angular radius 
(typical of the microlensing field) not angular diameter 
(conventional in the field of angular-size determinations).
We convert our derived color of the source $(V-I)_{\rm S,0}=1.07 \pm 0.05$,  
to $(V-K)_{\rm S,0}=2.40 \pm 0.15$ 
using the calibrations of Bertelli \etal\ (1994) (ATLAS9 models) and 
Bessel \etal\ (1998) (ATLAS9 + NMARCS models). 
These calibrations are nearly independent of metallicity and surface gravity. 
The surface brightness-color relation then yields an angular radius of 
the source of  
\begin{equation}
\theta_* = (5.56 \pm 0.54)\,\mu{\rm as~}.
\label{eqn:thetastar}
\end{equation}
The error in equation (\ref{eqn:thetastar}) assumes an intrinsic scatter 
of 8.7\% in the van Belle (1999) surface brightness-color relation, 
which we judge from the excess in the scatter over measurement uncertainties, 
and includes the 2.0\% standard error about the mean of the restricted 
relationship we have used as well as our measurement uncertainties in 
$I$, $(V-I)$, and $A_I$ of 0.02, 0.03 and 0.08~mag, respectively.  
Even rather large ($\sim$0.5~mag), unrecognized errors in the extinction  
would have a small effect ($ < 3.5$\%) on our estimate of $\theta_*$ 
because extinction is accompanied by reddening which affects the angular 
size estimate in the opposite sense.  

We stress that the angular size can be determined much more accurately 
than the physical size (or distance) of the source.  
We have checked our determination based on van Belle's (1999) empirical 
relation with angular sizes predicted from the isochrones of Bertelli \etal\ (1994).  We find that for a range of metallicities ($-0.4\leq$[Fe/H]$\leq 0$)
and ages (8 Gyr $\leq T\leq$ 12 Gyr) that reproduce the intrinsic color 
$(V-I)_{\rm S,0}=1.07$ of the source, stellar radii and distances 
(inferred from our $I_{\rm S,0}=14.67$) vary by almost
a factor of two.  Despite this, the angular radii formed by the ratio of 
physical radius and distance are extremely stable, within 
2\% of our slightly modified van~Belle (1999) relation.  We conclude that
equation (\ref{eqn:thetastar}) is robust.  Our estimate of $\theta_*$ 
is considerably larger than the $(2.9 \pm 0.7) \mu$as of 
Bennett \etal\ (1999); the cause of this is not completely clear. 
  
Combining our measurements for $\theta_*$ and $t_*$ yields
\begin{equation}
\mu ={\theta_*\over t_*} = (50 \pm 5)\,\kms\,\kpc^{-1},
\label{eqn:mueval}
\end{equation}
which corresponds to $(400\pm 40)\,\kms$ projected onto the Galactic center
at 8~kpc.  This is somewhat higher than would be expected for a bulge
source lensed by either a bulge or disk lens, although it could be consistent
with either.  This large proper motion is what one would expect 
for a disk lens lensing a disk source on the far side
of the bulge.  Since the source is close to the Galactic plane 
$(l,b)=(1.32,-1.94)$, such a geometry is not implausible.

%---------------------------------------------------------------------------

\subsection{Mass-Distance Relation}

%---------------------------------------------------------------------------

The Einstein ring radius can be estimated from our model and the 
proper motion estimate above as 
\begin{equation}
\theta_{\rm E} = \mu \, t_{\rm E} = (700 \pm 70)\,\mu\rm as.
\label{eqn:thetaeeval}
\end{equation}
Hence, from equation~(\ref{eqn:thetae}), 
we obtain the mass-source distance relation
\begin{equation}
\biggl({M\over M_\odot}\biggr)
\biggl({1-x\over x}\biggr)
\biggl({\dos\over 8\,\kpc}\biggr)^{-1}
= 0.47 \pm 0.10,\qquad {\rm ~~~~where~~}x\equiv {\dol\over\dos}. 
\label{eqn:massdis}
\end{equation}

If the lens were luminous and in the bulge, 
the total mass would be constrained
by the mass of the turnoff to be $M = (1+q)M_1 \la 1.5\,M_\odot$.
Hence, equation (\ref{eqn:massdis}) implies $x\la 0.75$.  Since lens-source
separations are distributed $\sim [1 + (1-x)^2/4(l^2+b^2)]^{-1}$
where $2(l^2+b^2)^{1/2}\sim 0.08$ (Gould 2000), 
this tends to argue against the lens being in the bulge.  We make this
argument more quantitative in the next section.

%---------------------------------------------------------------------------

\subsection{Statistical Estimate of Binary Lens Mass, Distance and Rotational Period}\label{distmassperiod}

%---------------------------------------------------------------------------

From the measurements of $d_{\rm mid}$, $\delta d$, $\delta \theta$, and
$\theta_{\rm E}$, we can determine the projected separation $r_p$ 
and the projected relative velocity $\vec v_p$ 
of the binary up to a scale set by the lens distance $D_{\rm L}$.
We use coordinates parallel and transverse to the separation vector $d_{mid}$ 
and consider only leading-order terms in $\delta d$ and $\delta \theta$. 
One then obtains $r_p = d_{mid} \, \theta_E D_{\rm L}$ and 
$\vec v_p = 
(\delta d/\delta t,\, d_{mid}\, \delta \theta/\delta t) \, \theta_E D_{\rm L}$. 
Inserting the parameter values determined from our model, we then find 
\begin{equation}
r_p = 1.4\, {\rm AU} \, \biggl({\dol\over 4\,\kpc}\biggr),\qquad
\vec v_p= (-9.7,6.9)\,\kms\biggl({\dol\over 4\,\kpc}\biggr)~.
\label{eqn:projmotion}
\end{equation}

These measurements can be combined with Kepler's Third Law to estimate
the mass and distance of the lens.  To do so, we assume that binaries of
the observed mass ratio $q = 0.34$ are distributed 
along the line of sight in proportion to the density of their primaries, 
that they have a period
distribution as measured by Duquennoy \& Mayor (1991) for G stars in the
solar neighborhood, that they are randomly oriented, and that they are
uniformly distributed in $e^2$, where $e$ is the eccentricity.  We
assume that the disk population has an exponential scale length of 3~kpc,
an exponential scale height that varies linearly from 300~pc in the
solar neighborhood to 100~pc near the Galactic center, and a local surface
density in stars and brown dwarfs of $35\,M_\odot\,{\rm pc}^{-2}$.  
We further assume 
that the bulge can be approximated as an isothermal sphere with a
rotation speed of $220\,\kms$, truncated at 4~kpc from its center.
We take the mass function to be $d N/d M \propto M^{-1}$ for both the disk
and bulge, but with a cutoff of $M_1\leq 1.1 \, M_\odot$ for the bulge.

We perform a Monte Carlo integration in order to find the relative  
probability densities of the mass, distance, and period of the binary.
First, we construct a set of random realizations of the orientation,
eccentricity, and eccentric anomaly $\psi$ (Goldstein 1980) 
of the binary.  For each
realization, we determine three quantities: $w_r\equiv r_p/a$,  
$w_v^2\equiv v_p^2/\langle v^2\rangle$, and
$\cos\Phi\equiv {\bf r}_p\cdot{\bf v}_p/(r_p v_p)$,
where $a$ is the semi-major axis of the
binary and $\langle v^2\rangle$ is its time-averaged square velocity.
We first ask whether $\cos\Phi$ is near (\ie\ within 10\%) 
of the observed value of 
$\delta d/[(\delta d)^2+ (d_{\rm mid} \, \delta\theta)^2]^{1/2}= - 0.81$.  
If it is not, we skip to the next realization.  
If it is, we use Kepler's Law and the Virial Theorem, 
$GM= a\langle v^2\rangle$, to find the period consistent with the observed parameters
\begin{equation}
P = 2\pi\,{w_v\over w_r}{d_{\rm mid} \, \delta t 
\over[(\delta d)^2+(d_{\rm mid} \, \delta \theta)^2]^{1/2}}
 = 3.59{w_v\over w_r}\,{\rm yr}.
\label{eqn:peval}
\end{equation}

Next, by eliminating $M$ in the Virial Theorem in favor of $\theta_E$, 
the value of which is fixed by equation~(\ref{eqn:thetaeeval}), we obtain 
\begin{equation}
x^2(1-x) = w_r w_v^2\biggl({c \delta t\over 2 \dos}\biggr)^2
\biggl(\theta_{\rm E} \, d_{\rm mid} \, [ (\delta d)^2 + 
(d_{\rm mid} \, \delta\theta)^2]\biggr)^{-1} = 0.265 \, w_r w_v^2,
\label{eqn:xeval}
\end{equation}
where we have assumed $\dos=8\,\kpc$ in the evaluation.
If $w_r w_v^2 >0.559$, this equation has no solution and we skip to the
next realization.  Otherwise, we find the two solutions and the corresponding
masses from equation~(\ref{eqn:massdis}).  If the higher-mass solution
has $M>1.9\,M_\odot$ we eliminate it because the lens would have then
given rise to detectable blended light 
(which is not observed: see \S\ref{rotbin} and Table~1).  
For bulge lenses we eliminate stars above the
turnoff, $M>1.5 M_\odot$ (\ie\ $M_1>1.1 M_\odot$).
Finally, we weight each realization
by the following factors: 1) $d t/d\psi= 1-e\cos\psi$ to take account
of the amount of time spent at each phase $\psi$, 2) the density of lenses
at $x$, 3) two factors of $x$, one for the cross section for lensing
($\propto x\theta_{\rm E}$) and one for the 
transverse flux of lenses relative to the sources ($\propto x\mu$),
4) factors to account for the mass function and period distribution as 
described above, 
and 5) factors to account for the transformation from the theoretical
quantities $(x,M,P)$ to the observables 
$(d_{\rm mid}\, , \delta d, \delta \theta)$.  
The most important of these last factors is that for $x$, which scales 
$\propto |x(1-x)/(2-3 x)|$, diverging at $x\sim 2/3$ thus compensating 
for the small number of simulated events near this value.  
Note that the factors in (3) are actually $\propto \dol \thetae$,
but since $\thetae$ is measured and $\dos \sim 8\,$kpc 
is approximately known, they can be taken as $\propto x$. 

\begin{figure}

\vglue -1.25cm

% COLOR FIGURE AND CAPTION (Not in color, just B&W version available)
\plotone{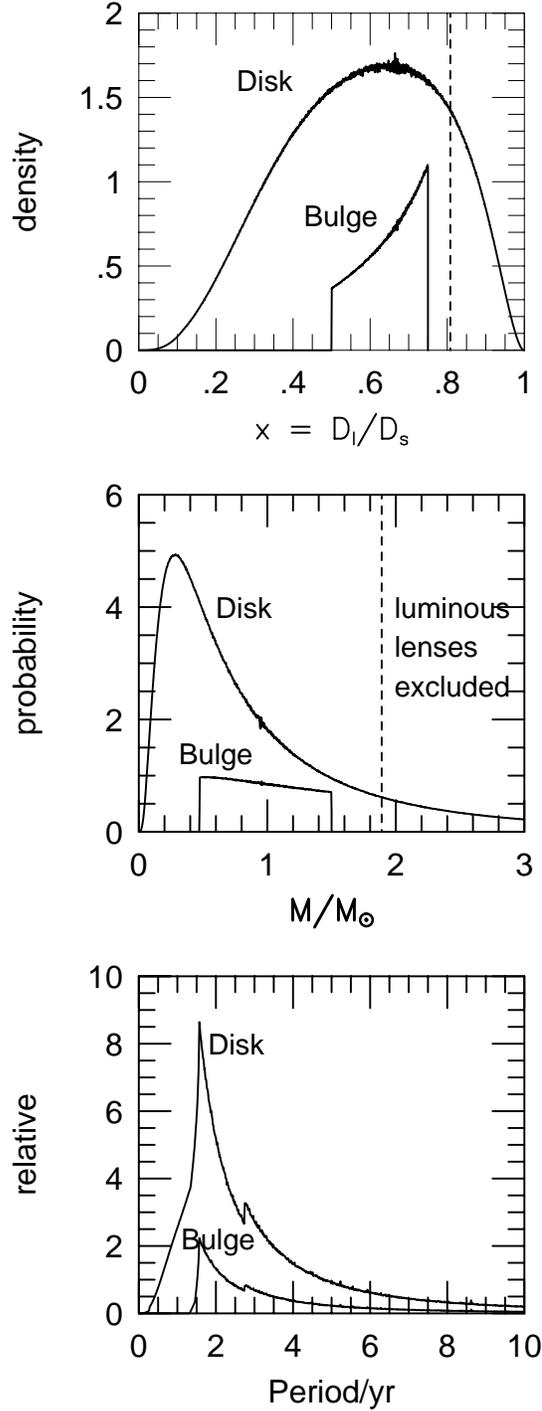}
\vglue -0.75cm
\caption[]{\label{fig:distmassperiod}
Physical properties of the binary derived from the rotation measurement and 
other observational constraints (see \S\ref{distmassperiod}).  
{\it Top:\/}
Relative probability density of lens distances $D_{\rm L}$ expressed as 
fraction $x$ of the source distance $D_{\rm S}$.  
The upper and lower lines are for disk and bulge lenses, respectively.
{\it Middle:\/} Same for the total mass of the binary lens.
{\it Bottom:\/} Same for the orbital period of the binary.
To the right of the dashed vertical lines 
luminous (stellar) lenses are excluded, 
since their mass would then correspond to more blended light than is observed.  
Bulge lenses have a lower distance cutoff (imposed by the simplistic 
bulge model) and an upper mass cutoff (imposed by the mass of bulge turnoff 
stars) which translate 
into mass and distance limits, respectively, 
through eq.~(\ref{eqn:massdis}). 
}

\end{figure}

Figure~\ref{fig:distmassperiod} shows the
results of this Monte Carlo integration for the lens distance, mass and 
orbital period, respectively.  
The noise near $x=2/3$ and $M\sim 0.95\,M_\odot$ is due to the low sampling
rate and corresponding large transformation factor described above.  
The probability is five times greater that the lens is in the disk than 
in the bulge.  Note that the peaks of the probability distributions 
for the lens mass and distance do not correspond to the same solution.  
If the lens is in the disk, the probability distribution for its total mass  
peaks at that of an M-dwarf binary with $M\sim 0.3\,M_\odot$, 
corresponding (from eq.\ \ref{eqn:massdis}) 
to a lens distance of $\dol \sim 3.1\,$kpc for a source distance of 
$\dos \sim 8\,$kpc.  
The peak at $\dol \sim 5.5\,$kpc of the probability distribution 
for the lens distance, on the other hand, 
corresponds to a total mass $M\sim 1.03\,M_\odot$.  
The probability distribution for the orbital period 
is rather sharply peaked around 1.5~yr.  
A firm detection of acceleration in the relative motion of the 
lens components would tighten the constraints on mass, distance and 
orbital period beyond that made possible by our detection of their 
relative velocity.

In order for an orbit to be bound, the ratio $r \, v^2/2\, G M$ must 
be less than unity. 
Solutions derived from our Monte Carlo procedure are bound by 
construction.  It can be shown that for any such solution,  
equation~(\ref{eqn:xeval}), together with the Virial Theorem 
and the definitions of $w_r$ and $w_v$, yields a constraint on 
this ratio for face-on orbits, namely
\begin{equation}
{{r_p \, v_p^2} \over {2\, G M}} = {{x^2 \, (1 - x)} \over {0.53}}~~.
\label{eqn:bounded}
\end{equation}
Since the right-hand side has an upper bound of 0.28 at $x = 2/3$, 
all such orbits are quite comfortably bound for any value of $x$. 
Only orbital phases and inclinations that would increase this ratio 
by a factor of four would result in unbound orbits.

%---------------------------------------------------------------------------

\section{Conclusions}\label{conclude} 

%---------------------------------------------------------------------------

We have presented our photometric data set for the complex 
\event\ event, consisting of 325 points in the $I$ and 46 points in the $V$ 
photometric passbands taken during the four month period in 
which the event was in progress.  The light curve displays two 
anomalous regions that cannot be fit by a static binary lens, 
in agreement with a similar conclusion by Bennett \etal\ (1999) 
modeling an independent data set.  Unlike Bennett \etal\ (1999), however, 
we find that a rotating binary provides an excellent fit to our data. 
We measure changes in the binary separation 
(in units of the Einstein ring radius) of $\delta d = -0.070 \pm 0.009$ 
and in the binary orientation on the sky of  
$\delta\theta = 5^\circ\hskip-2pt .61\pm 0^\circ\hskip-2pt .36$ 
during the 35.17 days between 
the separate caustic events, 
the first clear observational evidence for rotation in a binary microlens.

Our success in finding this rotating solution was facilitated by 
the high quality of our data for this event and by three strategic 
choices: (1) searching for a solution 
involving only the first-order rectilinear component of the relative 
motion of the binary on the sky, 
(2) fitting an empirical parameter set  
that is best constrained by the light curve rather than the equivalent, 
but more non-linear canonical parameter set, and 
(3) searching on a grid 
of solutions until a gross minimum is found rather than relying on 
downhill simplex methods only to search for approximate minima.

Bennett \etal\ (1999) used the MACHO/GMAN data set to reach 
their conclusion that a three-lens system (binary + Jovian planet) was 
indicated in \event. 
In contrast, we find that our rotating binary model provides a good fit 
to the same data with fewer additional degrees of freedom, 
even though the MACHO/GMAN data were not used in any way to fix 
the physical binary parameters and include points in  
caustic regions that our data do not sample well.  
Furthermore, the particular three-body model presented by Bennett \etal\ 
(1999) is strongly inconsistent with our data near the first anomaly 
in the light curve of \event.
We conclude that a third body is not required, 
and that binary lens rotation provides a natural, physically-motivated and 
physically plausible explanation of the light curve of \event. 

Our modeling also produces a limb-darkening parameter for the background 
K giant corresponding to $c_I = 0.52 \pm 0.10$, in agreement with models of 
stellar atmospheres for this type of star.  
Combining our physical and kinematic measurements with 
reasonable assumptions for the distributions in space and separation 
of Galactic binaries, we are able for the first time 
to derive kinematic probability distributions for the total mass and 
period of a binary microlens; for \event, these peak at 
$0.3 \, \msolar$ and 1.5~yr, respectively.

%---------------------------------------------------------------------------

\acknowledgments

\subsection*{Acknowledgments}

%---------------------------------------------------------------------------

PLANET thanks the MACHO collaboration for providing the 
original electronic alert of this event; such alerts are 
crucial to the success of our intensive microlensing monitoring. 
We also thank Andy Becker for providing us with the MACHO/GMAN data 
for this event, and Andy Becker and Dave Bennett for useful comments 
about photometric uncertainties in those data.
PLANET is grateful to Christophe Alard for advice in the use of ISIS, 
to Piotr Popowski for pointing out that estimates for the color and 
magnitude of the red clump had been revised, and to Gerard van~Belle for 
sending us data in advance of publication. 
We are especially grateful to the observatories that support our science 
(Canopus, CTIO, ESO, Perth, and SAAO) through generous time allocations, 
and to those who have donated their personal 
time to observe for PLANET at Canopus Observatory, 
including Paul Cieslik, Duncan Galloway, Josh Martin and John Phillips, 
as well as Bob Coghlan and several other members of the 
Astronomical Society of Tasmania.   
PLANET acknowledges financial support via award GBE~614-21-009  
from the organization for 
{\sl Nederlands Wetenschappelijk Onderzoek\/} 
(Dutch Scientific Research), 
the Marie Curie Fellowship ERBFMBICT972457 from the European Union, 
a ``coup de pouce 1999'' award from the 
{\sl Minist\`ere de l'\'Education nationale, de la Recherche et de la Technologie, D\'epartement Terre-Univers-Environnement\/}, 
grants AST~97-27520 and AST~95-30619 from the NSF, and NASA grant 
NAG5-7589.

\newpage

%---------------------------------------------------------------------------

\newpage

%---------------------------------------------------------------------------

% Tables

%---------------------------------------------------------------------------

% Table 1: Photometric Fit Parameters

\begin{deluxetable}{l c c c c c}
\tablewidth{14cm}
\tablenum{1} \label{photofittable}
\tablecaption{Photometric Parameters fitted to Cleaned PLANET Data for \event}
\tablehead{
\multicolumn{1}{c}{Data Set} &
\multicolumn{1}{c}{\# Points} &
\multicolumn{1}{c}{$\sigma/\sigma_{\rm ISIS}$} &
\multicolumn{1}{c}{$\etas '$\tablenotemark{a}} & 
\multicolumn{1}{c}{$I_S$\tablenotemark{b}} & \multicolumn{1}{c}{$I_B$\tablenotemark{b}} 
}
\startdata
~~~~~~~I~~BAND&     &      &           &      &      \\
SAAO 1m       &  97 & 1.15 & $-$0.0270 & 16.82 & 20.12 \\
ESO/Dutch 0.9m&  58 & 1.59 & $-$0.0178 & 16.82 & 20.33 \\
Canopus 1m    &  95 & 0.73 & $-$0.0632 & 16.81 & 19.44 \\
CTIO 0.9m     &  49 & 1.55 & $-$0.0290 & 16.89 & 21.57 \\
Perth 0.6m    &  26 & 1.21 & $-$0.0141 & 16.81 & 19.00 \\
              &     &      &           &      &      \\
~~~~~~~V~~BAND&     &      &           &      &       \\
SAAO 1m       &  14 & 1.45 & ~~0.2725  & 19.33 & 22.84 \\
ESO/Dutch 0.9m&  18 & 0.92 & $-$0.1304 & 19.38 & 22.03 \\
Canopus 1m    &  14 & 0.34 & $-$0.1508 & 19.43 & 19.56 \\
\enddata
\tablenotetext{a}{The parameter $\etas '$ is related to the seeing 
correlation parameter $\etas$ of \S\ref{model} by 
$\etas ' = \etas/(\fs + \fb)$, and thus has units of arcsec$^{-1}$.}
\tablenotetext{b}{Fluxes $\fs$ and $\fb$ have been converted to standard 
Johnson V and Cousins I magnitudes.}
\end{deluxetable}
\bigskip

%---------------------------------------------------------------------------

% Table 2: Geometric Fit Parameters

\begin{deluxetable}{c c c c c }
\tablewidth{12cm}
\tablenum{2} \label{geofittable}
\tablecaption{Geometric Parameters fitted to Cleaned PLANET Data for \event}
\tablehead{
\multicolumn{2}{c}{Empirical} & &
\multicolumn{2}{c}{Canonical\tablenotemark{a}}
}
\startdata
$d_{mid}$       & 0.5122 (0.0015)    &~~~~~~~~~& $d_0$       & 0.4787      \\
$q$	        & 0.3434 (0.0065)          &~~~& $q$	     & 0.3434         \\
$t_{\rm cc,1}$	& ~~~~619.18 (0.03) days   &~~~& $\te$       & 24.175 days \\
$t_{\rm ce,2}$	& ~~~~654.544 (0.001) days &~~~& $t_0$       & 653.4254 days\\
$u_{\rm c,1}$	& ~0.0020 (0.0005)~        &~~~& $u_0$       & 0.07229      \\
$\ell_2$	& 0.028 (0.001)	           &~~~& $\alpha$    & 112.37       \\
$\Delta t_2$	& ~~~~~~0.193 (0.005) days &~~~& $\rho_*$    & 0.00796      \\
$\delta d$	& $-$0.070 (0.009)~~~      &~~~& $\Delta_x$  & $-$0.0467    \\
$\delta \theta$ & $5^\circ\hskip-2pt .61 \, (0^\circ\hskip-2pt .36)$   &~~~& $\Delta_y$  & 0.0366         \\
$\Gamma_I$ 	& 0.42\tablenotemark{b} (0.09) &~~~& $c_I$	 & 0.455\tablenotemark{b}         \\
\enddata
\tablenotetext{a}{Precision given in canonical parameters is necessary 
to reproduce the model, but the last two digits are generally uncertain.}
\tablenotetext{b}{For the empirical parameters, $\Gamma_I$ corresponds to  
$c_I = 0.52 \pm 0.10$; for the canonical parameters, $c_I$ corresponds to 
$\Gamma_I = 0.358$.  For a discussion, see \S4.2.2.} 
\end{deluxetable}
\bigskip

%---------------------------------------------------------------------------

% Figures

%---------------------------------------------------------------------------

\newpage

%--------------- Figure 1 of 7 ---------------------------------------------

%--------------- Figure 2 of 7 ---------------------------------------------

%--------------- Figure 3 of 7 ---------------------------------------------

\newpage

%--------------- Figure 4 of 7 ---------------------------------------------

%--------------- Figure 5 of 7 ---------------------------------------------

%--------------- Figure 6 of 7 ---------------------------------------------

%--------------- Figure 7 of 7 ---------------------------------------------

\end{document}